# Information and Communication Technology in Combating Counterfeit Drugs

Haruna Isah

latuji@computer.org

Department of Computing, School of Informatics, Computing and Media (SCIM)

University of Bradford, Richmond Road BD7 1DP, Bradford UK

# Abstract

Pharma frauds are on the rise, counterfeit drugs are giving sleepless nights to patients, pharmaceutical companies and governments. The laws prohibiting the sales of counterfeit drugs cannot succeed without technological interventions. Several analytical techniques and tools including spectroscopy, holograms, barcoding, differentiated packing, radio frequency identification, fingerprints, hyperspectral imaging etc. have been employed over the years in combating this menace; however this challenge is becoming more sophisticated with the evolution of the World Wide Web and online pharmacies. This paper presents a review on the contribution of Information and Communication Technology (ICT) as a drug counterfeit countermeasure.

# Keywords

Cloud, Counterfeit, ICT, Drugs, HP, mPedigree, Pharmacy, Sproxil, WHO

# 1. Introduction

Pharmaceutical products which are produced and sold with the intent to deceptively represent its origin, authenticity or effectiveness are generally referred to as counterfeit medications [1]. This definition is in concert with the World Health Organisation's (WHO) definition; a counterfeit medicine is one which is deliberately and fraudulently mislabelled with respect to identity and/or source. The risk of fake drugs can lead to unexpected side effects, allergic reactions, worsening of medical conditions and even deaths; pharmaceutical companies and manufacturers are faced with brand issues and financial losses due to patent and copyright infringement, while the government in such areas suffer huge economic loss. Counterfeits make up more than 10% of the global medicines market and it has been estimated that 25% of the medicines consumed in poor countries are counterfeit or substandard. The sale of counterfeit drugs kills up to 10,000 Africans each year and creates 2 to 5% losses for governments around the world [2, 11].

The fight against counterfeit medication and drugs trafficking necessitates the collaboration of various disciplines, technologies and infrastructures that focuses on the legislation, regulation, enforcement and implementation of policies geared towards combating this menace. The major areas to be addressed for a successful fight against counterfeit medications as identified by [2, 11] includes; securing the drug product and its packaging, tracking product movement, enhanced regulatory oversight and enforcement, increased penalties, increased awareness of counterfeit drugs and international collaborations. The internet has evolved into a virtual market place to sale everything as such opening opportunities for illegal transactions such as the marketing and sales of counterfeit, unprescribed, new and unapproved drugs. There are few initiatives to increase the traceability of drugs and improve supply chain security and authentication; however, different pharmaceutical companies' uses different security features to distinguish the original from the fake drugs. The aim of this paper is to review the contributions of information and communication technologies in combating this menace.

## 2. Literature Review

Some of the techniques and strategies employed in combating pharma fraud includes: Radio Frequency Identification and Detection (RFID), Near Infrared Spectroscopy, Raman Spectroscopy, Fluorescence and Phosphorescence measurements, Nuclear Magnetic Resonance imaging, X-ray and radio frequency analysis, tamper resistant tape, fingerprints, holograms and colour-shifting inks and dyes etc. [3].

Recent research work on techniques for identifying and detecting fake drugs include the study by [4, 5, 6] on the automatic classification of counterfeit samples (Viagra, Cialis and Levitra) tablets using image processing and statistical analysis, optical spectrophotometer (colorimeter) and X-ray fluorescence (XRF) respectively. In [4] the detection was performed by computation and checking of the adherence of a test sample to the obtained distribution of colour pixels with reference to that of genuine tablets, also in [5] the colorimeter was tested on both original and counterfeited samples by measuring the colour of tablets' surface and of a specific spot of the packages; suspicious samples were then compared to the corresponding original (spectral libraries) by means of a wavelength distance pattern recognition method. In [6] a relatively simple forensic investigations or analytical methodology for semi-quantitative determination of active ingredients and excipients to classify tablets investigated between authentic and counterfeit groupings was presented. Another effective counterfeit drug detection technique investigated by [7] is the information-rich Near Infrared (NIR) spectrometry (using The Unscrambler 8.0 software) and multivariate hyperspectral image analysis (with a multivariate image analysis tool JIMIA, written in Java (by J. Burger), calibration and chemometric routines were coded in MATLAB). According to [7] Visual control, dissociating tests or simple colour reaction tests reveal only very rough forgeries hence the inclusion of the mathematical data processing technique. Other product and packaging assurance and evaluation method includes surface analysis which aids in optimisation and acceleration of new product development, evaluation of product and packaging stability, rapid identification of trace contamination and quality assessment of new manufacturing processes [8]. Radio Frequency Identification (RFID) is a track and trace technology which is intended to tag products by manufacturers, wholesalers and retailers hence reducing the risk of interference; however this technology is limited where products can be re-packaged. Again bar codes (huge blocks of numbers) stored on a central database is useful in authenticating batch processing and distribution [9].

## 3. ICT Tools and Techniques for combating counterfeit drugs

## 3. 1. Overview

Counterfeit drugs seep into the market mostly through retailers. For high turnover, counterfeiters target the most lucrative markets with weak or no fraud regulation and high demand for medicines. Most of the countermeasures reviewed earlier have either still remained untapped or failed. Information and communication tools largely find significance in counterfeit authentication and verification related activities, internet pharmacies regulations, organisational and stakeholder collaborations as well as in public awareness campaigns. Recent developments in combating pharma fraud with communication and web technologies include the cloud, mobile and syndication technologies, rapid alert systems, coded stickers, website seals and cyber warnings.

## 3. 2. Cloud Technology

Cloud technologies ensure that scalable and elastic information technology capabilities are provided as a service to multiple customers using internet technologies. Enterprises now act as cloud providers to deliver applications, information or business processes and services to customers and business partners. The system as employed in pharma fraud detection provides a counterfeit countermeasure services through Mobile Product Authentication (MPA). Geared towards emerging markets, the technology comprises of a hosting infrastructure and a business interface to check and verify the authenticity of medications. Once a firm registers its products, its drug packs will have the code, instructions and the number to which the short message services (SMS) should be sent.

The user scratches out and send the code on the strip of the medicine to the number provided by the firm and the message goes to the cloud server where the code is verified according to the data given by the pharmaceutical companies. The server then sends back an SMS on whether the medicine is authentic or counterfeit. The system also allows pharmaceutical companies to monitor the movement of products through their global supply chains with a much greater accuracy. This helps protect consumers against ineffective drugs and enables drug manufacturers to protect their revenue, brand and IP. It's a free service, funded largely by the pharmaceutical companies involved.

## 3. 2. 1. Sproxil

Sproxil uses IBM's cloud technology to provide clients with secure, reliable data access virtually anywhere, it allow consumers to verify the authenticity of prescriptions in seconds with their mobile phones. To make it easier for its clients to view and analyse this market data, Sproxil turned to IBM's ILOG Elixir software, which provides rich visuals such as advanced charts and graphics. Using these and other new capabilities, pharmaceutical manufacturers around the world will be able to better manage and analyse petabytes of transaction data in real time. Using IBM SmartCloud, Sproxil is benefiting from the cost savings and scalability associated with a cloud environment while preserving the ability to take advantage of the security, existing applications, reliability, management and support services more typical of a private cloud.

## 3. 2. 2. HP Labs and mPedigree Networks

Another Mobile Product Authentication solution utilizing the cloud technologies is mPedigree, where HP Labs runs the hosting infrastructure, security and the integrity of the service while the mPedigree Network provides the business process interface.

Another such similar system is the solution by [15] such that users' initiates counterfeit checking with a wireless device equipped with camera to snap a picture of anti-counterfeiting code pattern of a product item; and send the digital image to a host computer for a checking and verification of the product authenticity.

## 3. 3. Rapid Alert Systems and Product Recalls

Is a web-based system for tracking the activities of drug cheats, it provide the means where by the public who believe they have come across counterfeit drugs go online to report what they have discovered and attain the assessment of specialists. It was the initiative by the World Health Organization to increase public awareness, promotes rapid and transparent information sharing, and obtains commitment of governments and other sectors in addressing drug counterfeiting. The system aids in ensuring that only approved products are traded in the marketplace. Warning and recall notices are issued against products that do not meet regulatory requirements or those causing side effects.

## 3.4. Drug Product Information System

Drug Product Information System consist of a database of approved drug products and other medical related items or devices in a market place set up by a recognised regulating body. Prior approval from the regulating body must be obtained by drug merchants and retailers before a new product can be allowed to be traded in the market place. This often serves as a reference used for tracking, detection and reporting of counterfeit drugs.

## 3. 5. RSS Technology and the Social Web

Feeds from pharmaceutical agencies websites and other relevant regulatory agencies are aggregated (fetched, sorted and filtered) in a single view on desktop PC, through browser interface, via mobile devices or using web based feed readers. The technology makes it easy for the public, pharmaceutical companies and law enforcement agencies to keep to date with current trends globally from a single view. Social media tools, such as Facebook, twitter, Google+ etc. are also used for counterfeit advisories, information sharing, emergency warnings and alerts.

## 3. 6. Internet Pharmacy Regulations and Website Seals

Website Seals are images or objects displayed in response to public concern of safety on the websites of accredited Pharmacies having complied with the licensing and inspection requirements of an appropriate regulating body. The Verified Internet Pharmacy Practice Sites (VIPPS) is such typical seal developed by The National Association of Boards of Pharmacy (NABP) applicable to 8 districts (all 50 United States, the District of Columbia, Guam, Puerto Rico, the Virgin Islands, eight Canadian provinces, and New Zealand). Pharmacies displaying the VIPPS seal have demonstrated to NABP compliance with VIPPS criteria including patient rights to privacy, authentication and security of prescription orders, adherence to a recognized quality assurance policy, and provision of meaningful consultation between patients and pharmacists. VIPPS pharmacy sites are identified by the VIPPS hyperlink seal displayed on their Web site. By clicking on the seal, a visitor is linked to the NABP VIPPS site where verified information about the pharmacy is maintained by NABP [21].

The Medicines and Healthcare products Regulatory Agency (MHRA) is the UK government agency which is responsible for ensuring that medicines and medical devices work, and are acceptably safe. MHRA worked closely with Pfizer and other partners including Royal Pharmaceutical Society (RPS), HEART UK and The Patients Association. With a 24-hour counterfeit hotline for reporting and feedback, MHRA aims are being achieved through prevention, incident management and investigation. Through its operation and in conjunction with the Met Police, MHRA has succeeded in safeguarding the UK supply chain, taking down several non-complaint websites, seized millions of doses of medicines with the UK Boarder Agency, prosecuted the offenders and helped raise public awareness of the dangers of purchasing medicines online through TV series 'Fake Britain'. The major limitation of website seal is that sometimes vendors display the accredited seal even when they are not on the list and tracing such is always difficult.

## 3. 7. Coded stickers

A new initiative to protect against counterfeit medicine in Lebanon, marking the latest in a wave of measures introduced to combat the problem. The product comes in the form of a sticker that will be placed on boxes of medicine to help consumers identify counterfeit medicine.

The sticker, reading "Living Proof," contains a unique alphanumeric code, which patients can use to check the medicine's authenticity by calling a helpline. The codes are revealed once by scratching a panel on the box and cannot be reused after it is verified.

## 3. 8. Cyber Letter

These are warning letters sent electronically via the Internet to web sites whose online sales of prescription drugs may be illegal as well as informing them of the laws that govern prescription drug sales. This is typical of the U.S Food and Drug Administration. Cyber letter is not intended to be an all-inclusive review of the affected website or products but often a means for subsequent interaction between the regulators and the recipient of the letter to ensure compliance with the violated Act. [11]

## 4. Discussion

Counterfeits medicines are difficult to detect, investigate, quantify and regulate; fake drugs are easy to smuggle to countries with weak regulations or low standard of living (since it is usually cheap, attractive to the poor and provides employment and income opportunities).

Analytical techniques such as Radio Frequency Identification and Detection (RFID), Near Infrared Spectroscopy, Raman Spectroscopy, Fluorescence and Phosphorescence measurements, Nuclear Magnetic Resonance imaging, X-ray and radio frequency analysis, tamper resistant tape, fingerprints, holograms and colour-shifting inks and dyes etc. have been adopted in inspection and control of pharma products. Yet such techniques seem not to have mitigated counterfeiting to an appreciable dimension due to the sophistications caused by the evolution of the internet with it relative anonymity, the World Wide Web and online pharmaceutical practices.

Information and Communication Technology (ICT) is intervening by supplementing the analytical tools and techniques in combating this menace. The sections below highlights some successes achieved in combating pharma fraud through ICT.

| | |
|---|---|
| **Internet Pharmacy Regulations and Website Seals** <br> Medicines and Healthcare products Regulatory Agency (MHRA) UK, Pfizer, Royal Pharmaceutical Society (RPS), HEART UK and The Patients Association. <br> *This Crown copyright material is reproduced by permission of the Medicines and Healthcare products Regulatory Agency (MHRA) under delegated authority from the Controller of HMSO.* | Achievements following 2007 anti-counterfeiting strategy campaign: <br> ▪ Convicted 24 individuals <br> ▪ Secured sentences of: <br> ▪ 30 years and two months imprisonment <br> ▪ 5 years and 9 and half months suspended <br> ▪ 930 hours Community Service <br> ▪ £1,300 fines <br> ▪ Been awarded £66,000 costs <br> ▪ Secured Confiscation Orders of £15.3m under proceeds of crime legislation <br> ▪ Dedicated 24-hour counterfeit hotline: 020 3080 6701 or by email to counterfeit@mhra.gsi.gov.uk <br><br> Impact of Get Real, Get a Prescription campaign in collaboration with Pfizer: <br> ▪ 75% of those who were surveyed would think twice about where they purchase medicines <br> ▪ Counterfeit Hotline referrals in the two months following the campaign were more than double the referrals in any month prior <br> ▪ 1% of UK males surveyed in September 2008 expected rat poison to be found in counterfeit medicines; this figure increased to 32% in November 2009 following the campaign <br> ▪ 94% of all adults were exposed to the advertising campaign, on average 12 times. <br><br> Impacts of International Results (Operation Pangea IV in 2011): <br> ▪ Involved over 80 countries and over 160 agencies <br> ▪ Seized in the week a total of almost 2.6 million doses, valued at approximate £5 million <br> ▪ Over 13,500 websites taken down <br> ▪ Over 600 adverts removed from auction sites. <br> ▪ UK results (Operation Pangea IV in 2011) <br> ▪ Taking down of over 12,800 websites in conjunction with the Metropolitan Police Service. |
| | - Seizure of 1.5 million doses of medicine with the UK Border Agency.
- Reporting through national press, Primetime TV and Radio and TV series 'Fake Britain'.
- Impact of Operation Singapore:
- 72,000 packs of counterfeit medicine (over two million doses) with a retail value of £4.7m had penetrated the legitimate UK supply chain
- Seven batches of three medicines were the subject of a MHRA Class 1 recall.
- The MHRA seized 1.3m doses before they reached pharmacies.
- The counterfeit products contained between 50% - 80% of the corrective active pharmaceutical ingredient together with unknown impurities.
- The individual behind the UK part of this international conspiracy was prosecuted by the MHRA and sentenced to eight years' imprisonment.

Impact of Operation Valleta-Goldfinch:
- In 2005 a warrant was executed on Hickman's business address from which he claimed to operate a call center for an Indian company.
- In 2006 MHRA obtained a High Court injunction against Hickman to prevent Internet supply leading to three month sentence and £20,000 fine.
- In 2009 a MHRA prosecution leads to a guilty plea and a two year prison sentence.
- In 2012 a MHRA-NW RART (North West Regional Asset Recovery Team) joint financial investigation concluded with a £15.3m proceeds of crime assessment with an order to pay £14.4m.
- The MHRA remains committed to tackling the issue of counterfeit medical products and this remains a high priority for the Agency.

***This Crown copyright material is reproduced by permission of the Medicines and Healthcare products Regulatory Agency (MHRA) under delegated authority from the Controller of HMSO.*** |
| National Association of | To date, NABP has reviewed over 9,800 sites – only 3% of those online sites appear to |
| Boards of Pharmacy (NABP) Verified Internet Pharmacy Practice Sites (VIPPS) | be in compliance with pharmacy laws and practice standards.<br><br>VIPPS-accredited pharmacies meet nationally endorsed standards of pharmacy practice and are permitted to display VIPPS seal on their websites. Typical of such sites include Aetna Rx Home Delivery, Ameripharm, BioPlus, BioScrip, SUPERVALU, Walgreen, etc. |
| **Cyber Warnings**<br>FDA, NAFDAC, NABP, MHRA | Warning Letters have been sent in 2008 to 28 U.S. companies and 2 foreign individuals marketing a wide range of products fraudulently claiming to prevent and cure cancer.<br>Regulatory Action Against Ranbaxy's Paonta Sahib Plant in India etc. 2008 and 2009.<br>FDA Imposes Restrictions on Coast IRB due to Violations. 2009 |
| **Rapid Alert Systems (RAS)and Recalls**<br>WHO and most regulators | Global Public Awareness. Information and updates on product recalls, classification and reasons for the recall are available on most of the regulators' websites such FDA, NAFDAC, NAPD, MHRA. |
| **RSS Technology and the Social Web**<br>Available for Subscription on FDA, NABP NAFDAC etc. websites. | FDA, NAFDAC, NABP, Canada Health etc. provides RSS feeds for subscriptions and uses a number of social media tools (these includes; Buttons and Banners Toolbox, Facebook, Quick Responses (QR) Codes, mobile applications, social bookmarking, widgets, YouTube and the Video Gallery etc.) to share its content and provide access to reliable and timely health and safety information when, where and how you need it. The significance of these technologies cannot be overemphasised even though it will be difficult to quantify its effectiveness. |
| **Cloud Computing** | The recorded success for this one can be found in Sproxil and mPedigree websites. |

# 5. Conclusion

ICT is playing a major role in regulating and tracking the import/export and market surveillance of counterfeit drugs through the rapid alert systems, cyber warnings, website seals and supply chain monitoring. Authentication of medical products is now being carried out within minutes using a mobile product authentication solution (information decision support) that utilizes the cloud technology, a business network interface and the global system for mobile communication (GSM). The recorded success is high according to the service providers, and largely because; mobile phones are ubiquitous and extremely common in most of the susceptible areas (emerging markets). Mobile Product Authentication operates in countries where even electricity and internet are not guaranteed. The technology also collects a lot of data about how and when medication is being sold, which isn't easy to track in the developing world.

Syndication technology, databases of certified drug manufacturers/merchants and information networks are beginning to improve the efficiency and effectiveness of how information is being exchanged between regulatory authorities as well as in creating public awareness to drug quality and safety issues.

Although the recorded successes in combating drug counterfeits through ICT in recent years have been commendable, yet there are still challenges and hindrances to counterfeit fight which requires continued technological updates especially on issues having to do with how to tackle scalability, technology duplication and the tendencies counterfeit drug merchants utilizes to elude nominal safeguards.

# Acknowledgments

Special thanks to everyone who is putting efforts to ensure that pharma frauds in its entirety are eliminated.

**Author Profile**

Haruna Isah: a Graduate Assistant with the Electrical and Computer Engineering Department, Federal University of Technology Minna. He obtained a Bachelor of Engineering (B. Eng.) Degree, from the department of Electrical and Electronics Engineering, University of Maiduguri in 2008; Masters of Science (MSc.), Software Engineering in 2012 from the School of Computing Informatics and Media (SCIM) University of Bradford, United Kingdom. He has authored a book "Full Data Controlled Web Based Feed Aggregator" and published an article with same title in the International Journal of Computer Science & Information Technology (IJCSIT). He is also a Professional Member of the Association for Computing Machinery (ACM), IEEE and the Computer Society of IEEE. His research interest includes Software Re-use, Data mining, Web Technologies, the Social Web and Artificial Intelligence.

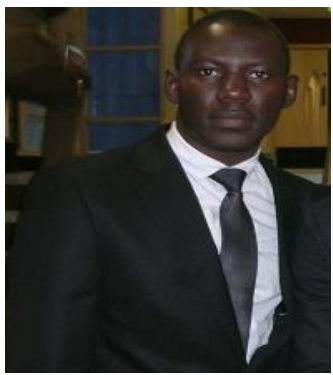